\documentclass[number,nonatbib]{elsarticle}
\usepackage{url}
\usepackage{verbatim}
\usepackage{graphicx}
\usepackage{cite}
\usepackage{adjustbox}
\usepackage{enumitem}
\usepackage{pgfplots}
\usepackage{tikz,amsmath,bm,color}
\usetikzlibrary{circuits.logic.US,circuits.logic.IEC,fit}

\usepackage{multirow}
\usepackage{pbalance}
\newcommand{\hide}[1]{}
\newcommand{\code}[1]{\textsf{#1}}
\newcommand{\TODO}[1]{}

\begin{document}

\title{Reproducing, Extending, and Analyzing\\ Naming Experiments\tnoteref{t1}}
\tnotetext[t1]{This research was supported by the ISRAEL SCIENCE FOUNDATION (grant no.\ 832/18).}


\author[1]{Rachel Alpern\fnref{fn1}}
\author[1]{Ido Lazer\fnref{fn1}}
\author[1]{Issar Tzachor\fnref{fn1}}
\author[1]{Hanit Hakim}
\author[1]{Sapir Weissbuch}
\author[1]{Dror G. Feitelson\corref{cor1}}

\fntext[fn1]{These authors contributed equally to the research}
\cortext[cor1]{Corresponding author: feit@cs.huji.ac.il}

\affiliation[1]{organization={The Hebrew University},
addressline={Dept.\ Computer Science},
postcode={91904},
city={Jerusalem},
country={Israel}}



\begin{abstract}
Naming is very important in software development, as names are often the only vehicle of meaning about what the code is intended to do.
A recent study on how developers choose names collected the names given by different developers for the same objects.
This enabled a study of these names' diversity and structure, and the construction of a model of how names are created.
We reproduce different parts of this study in three independent experiments.
Importantly, we employ methodological variations rather than striving of an exact replication.
When the same results are obtained this then boosts our confidence in their validity by demonstrating that they do not depend on the methodology.

Our results indeed corroborate those of the original study in terms of the diversity of names, the low probability of two developers choosing the same name, and the finding that experienced developers tend to use slightly longer names than inexperienced students.
We explain name diversity by performing a new analysis of the names, classifying the concepts represented in them as universal (agreed upon), alternative (reflecting divergent views on a topic), or optional (reflecting divergent opinions on whether to include this concept at all).
This classification enables new research directions concerning the considerations involved in naming decisions.
We also show that explicitly using the model proposed in the original study to guide naming leads to the creation of better names, whereas the simpler approach of just asking participants to use longer and more detailed names does not.
\end{abstract}





\maketitle
\thispagestyle{plain}

\section{Introduction}

Choosing variable names is one of the most common yet important tasks that programmers perform in their everyday workflow.
Variable and function names account for around a third of the tokens and two thirds of the characters in the source code of large open source projects \cite{deissenboeck06}.
It is generally agreed that meaningful names are instrumental aids for program comprehension \cite{brooksr83,gellenbeck91,blinman05,salviulo14}.
However, names can also be misleading and cause difficulties \cite{arnaoudova10,avidan17},
and low quality names have been associated with low quality code \cite{butler10}.
It is therefore important to understand the process of name selection, and its effect on the quality of names and code.

Previous research on naming largely falls into two categories.
Some studies are concerned with the more technical aspects of names.
This includes structural issues such as names' length or style \cite{butler15,butler15b,alsuhaibani21}.
There have been conflicting findings regarding the effect of using abbreviations on comprehension \cite{lawrie06,scanniello13,hofmeister17,schankin18}.
Several works have considered the extreme case of single-letter names, showing that even they may convey meaning \cite{beniamini17,swidan17}.
Conversely, longer names have been linked to variables with broader scope \cite{aman21,feitelson23b} and to developers with more experience \cite{feitelson22}.
Regarding style, several researchers have attempted to determine whether camelCase or snake\_case have any advantage \cite{binkley09b,sharif10,binkley13}.

Other studies focus on the semantics of names.
For example, several studies have analyzed the grammar and part-of-speech tags of words used in names \cite{newman20,newman:tags}.
Interesting observations have been made based on identification of renamings performed during program maintenance.
One common change is that names become more narrowly focused when words are added to a name \cite{peruma18}.
Alternatively, the semantics of names may actually change when words in a name are replaced \cite{arnaoudova14}.

There have also been suggestions on how to improve name quality.
For example, Caprile and Tonella suggest standardization of variable names using a lexicon of concepts and syntactic rules for arranging them \cite{caprile00}.
Deissenboeck and Pizka stress the need for concise and consistent naming \cite{deissenboeck06}.
Binkley et al.\ suggest rules for enhancing the information expressed by field names \cite{binkley11}.
The possibility of tool support for selecting names has also been discussed.
Liu et al.\ have shown that in many cases the contents of the body of a function can be used to construct a good function name \cite{liu19}.
Raychev used machine learning on ``big code'' to predict variable names, achieving 62\% accuracy \cite{raychev15}, and Alon et al.\ exploit the code structure to predict method names \cite{alon19a}.

But how do developers choose names in practice?
A recent study by Feitelson et al.\ included experiments where developers were presented with various scenarios, and asked to name objects that are expected to be used in those scenarios \cite{feitelson22}.
This study is unique in that the generated data shows how multiple developers would name the same objects --- and in particular, the variability that may result.
And indeed the study authors observed that different developers tend to choose different names under identical circumstances.
However, these names were nevertheless found to often include common structures and words.
In addition, a 3-step model was suggested for how names are created.

An important methodological challenge faced by this study was the danger of an accessibility bias.
By describing a scenario we might implant the words used in the description in the minds of the experiment participants.
It would then be natural for them to use these same words when naming objects.
To reduce this danger the study employed bilingual subjects, and presented the scenario in Hebrew to distance it from the writing of code.
A followup study used emojis in place of key words in the description, and achieved similar results \cite{regev21}.

The accessibility bias is just one of many threats to validity that the original study, like any empirical study, may suffer.
Such threats can be mitigated, at least to some degree, by using careful and innovative experimental designs --- such as using bilingual subjects.
But even research that seems to clear all known threats cannot be considered the final word on a subject.
There is always a possibility that some unknown limitation or interactions between the experimental variables affects the results.
The best way to validate results is therefore to reproduce them using alternative methodologies \cite{feitelson15}.
If the results are found to corroborate those of the original study, this will then increase the confidence that they are valid, because the same results were obtained by two different methodological approaches.

This leads to our main motivation and first research question:
\begin{enumerate}[label=RQ\arabic*), ref=RQ\arabic*, left= 0pt]
    \item\label{rq:reproduce}
    Can the results of Feitelson et al.\ \cite{feitelson22} be reproduced using variations on the experimental methodology?
\end{enumerate}
Our main variation is to replace the descriptions of the scenarios with actual code, but with meaningless variable names, and ask participants in the experiment to rename them and give them meaningful names.
This achieves two goals.
First, it is an alternative approach to reducing the accessibility bias.
Second, this alleviates an additional limitation of the original study, which was limited to the naming of central variables and data structures.
The reason was that it would be unreasonable to refer to specific local variables in a general description of a scenario.
But in our code-based experiments participants also need to name parameters and local variables.

We also extend the analysis of the names themselves in terms of their semantics.
The original study introduced a methodology for identifying the concepts embedded in a name and the words used to represent these concepts.
We use this methodology to dissect the variable names in our experiments, and show that their structure generally agrees with the findings of the original study.
But we also ask an additional question:
\begin{enumerate}[label=RQ\arabic*), ref=RQ\arabic*, left= 0pt, resume]
    \item\label{rq:concepts}
    Why do different developers come up with different names for the same objects?
\end{enumerate}
To answer this we define a classification of concepts according to the level of variance in their use.
We hypothesize that concepts that exhibit high variability expose disagreements in the naming process, and that this is an important source of naming variations.
This is important because it provides clues regarding issues that should be studied further, with the ultimate potential for drafting guidelines about the considerations that may be applied when choosing names.

Finally, the original study included an intervention experiment, where participants were instructed about the suggested model of name construction before being asked to give names.
The result was that names given by participants instructed about the model were generally considered better by external judges, and were also slightly longer.
Our third research question asks whether a simpler alternative intervention may suffice:
\begin{enumerate}[label=RQ\arabic*), ref=RQ\arabic*, left= 0pt, resume]
    \item\label{rq:better}
    Can higher-quality names be obtained by simply telling developers that longer names are better?
\end{enumerate}
We therefore included such a third treatment in our reproduction of this last experiment.
The result was that actually knowing about the model is required, and that telling participants that longer names are better is not enough to change their behavior.

The rest of this paper is structured as follows.
The next section describes the original study by Feitelson et al.\ \cite{feitelson22}.
Sections \ref{sect:exp1} and \ref{sect:exp2} then describe the first two experiments we conducted.
The methodology for name analysis is laid out in Section \ref{sect:concepts}, and the results presented in Section \ref{sect:res1}.
Section \ref{sect:exp3} then describes the intervention experiment, followed by Section \ref{sect:judge} which explains the methodology of judging names' quality and Section \ref{sect:res2} which presents the results.
The paper ends with threats to validity in Section \ref{sect:threats} and conclusions in Section \ref{sect:conc}.

\section{The Original Study}
\label{sect:orig}

To enable a clear understanding of how our reproduction differs from and extends the original study, we first describe that study \cite{feitelson22}.
The study was based on the definition of 11 independent scenarios.
Each scenario described a certain programming problem, followed by several questions related to a hypothetical program that could be written to address this problem.
Except in one scenario, there was no actual code --- just a verbal description.
The questions were about the naming of variables, constants, or functions in this hypothetical program, or about understanding such names which allegedly come from the program.
For example, one of the scenarios was of a mouse searching a maze for cheese which is moved every day, and two of the questions were about naming the data structure describing the maze and the variable which notes the location of the cheese today.
Eight of the scenarios had 2 or 3 versions, with the description of the scenario given in Hebrew and English.
Three additional scenarios had only a Hebrew version, and contained only name understanding questions.

In the first experiment in the study each participant was asked to answer 6 randomly chosen scenarios.
The data that was collected led to a plethora of results, including the following.
The comparison of names given for English and Hebrew versions of the same scenario was used to demonstrate the accessibility bias.
As expected, participants given the English version tended to use words from the description, in some cases to the exclusion of all else.
Participants given the Hebrew version created much more diverse names.
This effect was quantified using two metrics which we use too:
\begin{itemize}
    \item The \emph{focus}, defined as the fraction of times that the most popular name or word was used;
    \item The \emph{diversity}, defined as the quotient of distinct names or words divided by total names or words.
\end{itemize}
The diversity of the names was used to show that in most cases there is a very low probability that two developers would choose the same name in the same situation.
In addition, the results were used to study the distribution of name lengths.
This showed that experienced participants tend to use slightly longer names, with more concepts.
Sex, however, did not have any effect on name length.

In the present paper we reproduce the results concerning the partitioning of names into words, and the distributions of focus and diversity.
Then we extend the analysis of the concepts, and introduce a classification of concepts based on how they are used and how they relate to other concepts.

In the original study the quantitative results were followed by a qualitative discussion of name structure.
The names used in programs are often composed of multiple words.
The analysis indicated that these represent different concepts, and it is the combination of these concepts which bestows meaning to the name.
This led to the formulation of a 3-step model of name formation: first, select the concepts to include in the name; second, select words to represent the different concepts; and third, decide how to combine these words to create the name.

Following the derivation of this model, a second experiment was presented.
This second experiment was like the first, with the difference that the aforementioned model was used to guide participants in their naming.
The results were that using the model appeared to facilitate the creation of better names, as judged by external judges (that is, the judges did not know the details of the experiment).
In addition, the names produced in this second experiment tended to be slightly longer than those from the first experiment.
In the present paper we also reproduce this second experiment, and extend it with a third treatment, where participants are just told that longer names are better.
This enables us to distinguish between a possible improvement due to causing participants to focus more on the names, and a possible improvement specifically due to using the model.

We note that this study was also replicated recently by Mi et at., with an emphasis on naming by students \cite{mi22}.
In their case Chinese was used for describing the scenarios, as it was found to lead to a smaller effect on the results than English descriptions.
The results were largely the same as in the original study.
An interesting new result was that the improvement using the model is especially large for beginning students, who otherwise tend to use especially short names.
Our work differs from this replication due to our analysis of the names' structures and the use of a third treatment when assessing the effect of the naming model, as described above.

\section{Reproduction Experiment 1}
\label{sect:exp1}

As noted above, our reproduction started with the original naming experiment.
We designed and executed two independent naming experiments, described in this section and the next, using a new methodology;
having two experiments allows for additional methodological variations, which further supports the validity of the results (in case they agree).
We then performed a more detailed analysis of the names' structures than in the original study.
This will be described in section \ref{sect:concepts}.

\subsection{Experimental Materials}

Our approach for creating the experimental materials for this experiment was to use a short program with a well-defined purpose, and replace the meaningful variable names in it with generic ones (\code{a}, \code{b}, \code{c}, etc.).
We had to decide if we wanted to write the code ourselves or find something existing online.
We realized that finding suitable code may hard, because it needed to fulfill some very specific requirements: it had to be understandable without any comments or variable names, and it had to have a range of interesting variables.
We eventually decided on writing our own custom code in this experiment, and using real code in the second experiment.

Writing the code was challenging because we needed something understandable but not too simple, with a wide variety of variables, so that the naming task would be non-trivial.
We chose to write a rich description of our code, so that the variables could represent many different details that arise from the description.
After we wrote a program that had the complexity level we aimed for, we needed to avoid creating an accessibility bias.
We used two different methods.
First, we wrote the description in Hebrew, as was done in the original study.
While this can not eliminate all the bias, it means that the participants at least couldn’t use our words directly when naming the variables.
Second, we chose function names that are overly-descriptive.
This means that the participants can easily understand what the functions do, but cannot use the function names directly in naming the variables.

The code we wrote deals with a quiz published by a dairy company to promote its business%
\footnote{Unfortunately this choice may lead to confusion between make-believe  participants answering questions in this quiz and real participants answering questions in our experiment.
We consistently use ``quiz'' when referring to the code to avoid such confusion.}.
The code includes functions to read an input file with the answers of people who had answered the quiz, to score these answers based on a given scoring rubric, to select 3 winners out of those who had achieved the highest score, and to send them personalized email notifications.

\subsection{Experiment Structure}

The experiment had four main sections.
The first was a general introduction to the experiment.
It explained that this is an experiment about naming variables which is expected to take about 10 minutes, and that the code is in Python.
It then stressed that all collected data will be used for research only, that no identifying information is collected, and that continuing to the questionnaire indicates agreement to participate in the research.

The second section was the longest and constituted the experiment itself.
It started with a general description of the dairy company's quiz, written in Hebrew.
This was followed by 5 code segments: the definition of global constants, the main function, and 3 service routines.
Each segment was first introduced by a short Hebrew description.
The code itself was shown as an image with conventional color highlighting.
While the functions had excessively long descriptive names as explained above, the variable names seen by study participants were single letter names in alphabetical order, starting with the service routines and then the main function.
However, some letters were skipped, e.g.\ common loop index variables \code{i} and \code{j}.
The task was to give better names to all the variables in the code segment.

The third section came after the questions about the variables in all the code segments.
This was a short survey concerning the considerations used in the naming.
Considerations we asked about included keeping the names concise, making them distinct from each other, making the code understandable, etc.
Participants in the experiment were asked to note the importance of these considerations on a 5-point scale.

The final section of the experiment contained demographic questions about programming language use, educational background, experience, sex, and age.

\subsection{Experiment Execution}

The experiment was conducted using the Google Forms platform.
Participants were recruited via personal contacts and solicitation in our department's computer labs, aided by chocolate treats.
All the participants naturally knew Hebrew.
All told we had 71 participants in this experiment.
79\% were male and 18\% female; 3\% preferred not to identify.
The participants had a quite extensive range of (professional) programming experience.
A third, 33\%, had 1--2 years of experience.
Another 30\% had 3--5 years, and 11\% had 6--9 years.
Just over a quarter, 26\%, had 10 years experience or more.
The same wide distribution holds for education.
13\% were current students, 41\% had a first degree, and 31\% had an MSc.
The rest had learned to program at school or in the army (10\%) or were self-taught (6\%).

\section{Reproduction Experiment 2}
\label{sect:exp2}

The second experiment, in contradistinction to the first, used real code taken from GitHub.
It also obfuscates function names rather than using overly descriptive names.
This complements the first experiment and reduces the threat that our programming style affects the results.

\subsection{Experimental Materials}

In this experiment we used real code from open-source repositories on GitHub.
Using natural code as opposed to synthetic code is expected to give us a more realistic setting for our naming questions.
The considerations we applied when choosing the code samples were the following:
\begin{itemize}
    \item \emph{Programming language}: we used Python --- a simple, well known language, familiar to most programmers as well as to new students.
    \item \emph{Complexity}: we wanted to have simple functions, which have a clear goal, so participants will understand them correctly and be able to propose meaningful variable names.
    However, we needed to avoid trivial functions, in order to motivate participants to invest thought in finding variable names that convey meaningful intent.
    \item \emph{Length}: we chose relatively short functions, considering the fact that long functions may take more time to understand and may reduce the cooperation of the participants.
    \item \emph{Domain}: We chose to use code taken from utility libraries, to avoid any need for specific domain knowledge and any dependence on knowing the common jargon used in a specific domain.
    This again makes the experiment more accessible to all participants.
\end{itemize}

To apply these considerations, one of the authors initially selected 10 candidate code snippets from various utility libraries hosted on GitHub, and masked the variable names.
Then another author performed the task of assigning meaningful names to the variables, and at the same time assessed how easy each snippet was to understand so that it would not take too long during the experiment.

Based on the results of this second author we chose 3 code snippets for the experiment.
These snippets performed the following functions.
The first was taken from a Python lists utility library%
\footnote{\code{https://github.com/gilbertohasnofb/listools/blob/master/listools/listutils/ list\_mask.py}}.
This function receives two input lists, a data list and a masks list, and filters the first list using the indications from the second list.
The indications can be \code{0}/\code{1} or \code{True}/\code{False}.
Our version was slightly abbreviated: we removed the initial check that the inputs are indeed instances of \code{list}.

The second function comes from a Python string utilities library%
\footnote{\code{https://github.com/daveoncode/python-string-utils/blob/master/ string\_utils/genera\-tion.py\#L41}}.
This functions creates a random string of the requested length using upper and lowercase letters and digits.

The third function is also from this library%
\footnote{\code{https://github.com/daveoncode/python-string-utils/blob/master/ string\_utils/manipu\-lation.py\#L324}}.
This function converts a string in snake\_case to using camelCase style.
A flag parameter can be used to make the initial character lowercase, and another parameter allows the words separator in the input to be any string rather than the underscore character.
Our version was again slightly abbreviated, with input checks removed.

The original function names were all removed and replaced by \code{foo} to make respondents focus on understanding the code, and to avoid the use of words from the function names as a pool of candidates for use in variable names.
This approach is the opposite of the approach used in the first experiment as described above.
The variable names in the code were replaced by single-letter names in alphabetical order, starting anew from \code{A} in each function. 
Participants were asked to replace these arbitrary names with meaningful ones.

\subsection{Experiment Structure}

The experiment structure was similar to that of the first experiment.
The first section was a general introduction to the experiment.
It explained that this is an experiment about naming variables which is expected to take about 15 minutes.
It then stated that no identifying information is collected, and that continuing to the questionnaire indicates agreement to participate in the research.

The next 3 sections constituted the body of the experiment.
Each started by displaying the code of one of the functions, shown as an image with conventional color highlighting.
This was followed by questions soliciting better names for all the variables in the functions (including both parameters and local variables).

The final section of the experiment contained demographic questions about sex, age, experience, educational background, and programming language use.

\subsection{Experiment Execution}

The experiment was carried out using the Google Forms platform.
As we wanted to randomize the order of the 3 functions, we created 6 equivalent forms with all possible orders.
We created a separate load balancer which chose one of these versions at random, and used this in invitations to participate in the experiment.
\TODO{what was the balancer platform?}
In order to raise motivation among the participants, we randomly selected 10
participants who completed the survey to win a small prize.
\TODO{what was the prize? was this really done?}

Participants were recruited via personal contacts and solicitation in our department's computer labs.
This was hampered by reduced attendance due to the Corona pandemic, and despite our best efforts we managed to enlist only 45 participants in this experiment.
73\% were male, 22\% were female, and 4\% did not identify.
They had a roughly similar distribution of experience as the participants in the first experiment, with 41\% having 0--2 years of experience, 16\% having 3--5 years, 24\% having 6--9 years, and 19\% having 10 or more.
In this case 27\% were BSc students and another 16\% already had the degree, 22\% were MSc students and 11\% had an MSc, and 25\% did not have a formal CS education.

\section{Names Analysis}
\label{sect:concepts}

As noted above, an additional goal of our work --- apart from reproducing the experiment from the original study --- is to extend the analysis of the names selected by the experiment participants.
In particular, we focus on the names' structure: the words that compose them, and these words' meanings.

To enable this analysis, we start by partitioning each name into words.
This is done by a simple routine which identifies the camelCase and snake\_case styles of word composition.
To focus on the meaningful words, we ignore prepositions such as ``of'', ``and'', etc.\ (except in the statistics of names' lengths).

The next step is to correct obvious spelling errors in words used in names.
The justification for this is that we are interested in the meaning of the names and not in the English and typing skills of the participants.
However, we retained cases which can be interpreted as abbreviations, because these are intentional different representations of a concept.
For example, in \code{crct\_answr} the second word was left as it was and not corrected to \code{answer}.

The heart of our analysis, following the original study, is to identify the concepts included in each name and the words used to represent each concept.
This was done manually, with the aid of the interactive tool developed in the original study.
In this we diverge from the common practice of using statistical measures such as term entropy to study semantic effects \cite{arnaoudova10,posnett11,ray16}.
Manual identification facilitates a much more nuanced classification, including considerations for removing ambiguity, as demonstrated by some examples shown below.

The tool used supports the analysis of all the names given for each variable, one such variable at a time.
All the names chosen by the study participants for this variable are displayed in a column one beneath the other.
The analyst can then interactively add columns that represent different concepts he or she identifies in the names.
Words representing these concepts are copied to the appropriate column.
Whenever a new word is copied, the tool scans all subsequent names, and if the same word appears in them, it automatically copies it to the same column for those names too.
This reduces the manual effort of recording all the classifications.
The end result is a matrix with a row for each name, as exemplified in Figure \ref{fig:concept-tool}.
The first cell in the row contains the name itself.
Subsequent cells contain words in the name that represent the concepts represented by the different columns.
If a name does not include a certain concept, the corresponding cell is left empty.
This procedure was performed by one of the authors for all the variables in both experiments.
The resulting classification was then verified by another author.
Any changes made in this second pass were then approved by the first author.

\begin{figure}\centering
    {\small\begin{tabular}{@{}l|l@{~~}l@{~~}l@{~~}l@{~~}l@{~~}l@{}}
    \hline
    & \multicolumn{6}{c}{\emph{Concepts}} \\
    \emph{Name} & \emph{Score} & \emph{Question} & \emph{Answer} & \emph{Correct} & \emph{Per}  & \emph{Number} \\
    \hline
    \code{correct\_answer\_score} & score & & answer & correct & & \\
    \code{correct\_answer\_points} & points & & answer & correct & & \\
    \code{points\_per\_question} & points & question & & & per & \\
    \code{points} & points & & & & & \\
    \code{points\_for\_correct\_q} & points & q & & correct & for & \\
    \code{points\_for\_question} & points & question & & & for & \\
    \code{question\_val} & val & question & & & & \\
    \code{answer\_score} & score & & answer & & & \\
    \code{question\_score} & score & question & & & & \\
    \code{score\_per\_question} & score & question & & & per & \\
    \code{num\_points\_for\_correct\_answer} & points & & answer & correct & for & num \\
    \code{score} & score & & & & & \\
    \code{successful\_answer\_score} & score & & answer & successful & & \\
    \code{num\_points\_correct\_answer} & points & & answer & correct & & num \\
    \code{points\_per\_correct\_answer} & points & & answer & correct & per & \\
    \code{correct\_val} & val & & & correct & & \\
    \code{points\_per\_answer} & points & & answer & & per & \\
    \code{pts\_per\_ans} & pts & & ans & & per & \\
    \code{correct\_ans\_point} & point & & ans & correct & & \\
    \code{correct\_score} & score & & & correct & & \\
    \hline
    \end{tabular}}
    \caption{\label{fig:concept-tool}\sl
    Examples of names and their analysis into concepts, for the \code{GLOBAL\_C} constant in Experiment 1.
    This constant represents the points awarded for each correct answer in the quiz.
    The full results included many more names (including repetitions) and a few more rare concepts.}
\end{figure}

In analyzing concepts there were a few instances where we left words unclassified if they did not make sense (and in rare cases a whole name was wrong for the given variable).
There were also a few cases which required extra care.
For example, the code in Experiment 1 tabulating the score of the quiz copied each answer to a local variable and then cast it to an \code{int} in another variable.
This second variable was named \code{answer\_i} by several participants.
But by comparing with the first variable, it appears that the \code{i} could mean either \code{int} or \code{index} --- two different concepts.
An opposite example was found in Experiment 2.
The first function there had a list of Booleans as a parameter, used as indicators to filter another list.
Two of the names given to this parameter were \code{index\_map} and \code{items\_bool}.
Taken in isolation, ``index'' and ``item'' are different concepts: one indicates a location, the other what is found at that location.
Likewise, ``map'' is an operation on list elements and ``bool'' is a datatype of list elements.
However, we decided that in this case the names are actually semantically equivalent, and assigned both \code{index} and \code{items} to the concept ``list items'', and both \code{map} and \code{bool} to the concept ``filter''.

Extending a distinction already alluded to in the original study, we classify the concepts included in a name into the following categories:
\begin{itemize}
    \item \emph{Universal} --- a concept that is in the consensus, with all or nearly all names including a word representing this concept.
    Example: the concept ``score'' in Figure \ref{fig:concept-tool}.
    \item \emph{Correlated} --- one of a pair of concepts which tend to appear together, because their combination is a phrase that expresses the desired semantics.
    Example: the concepts ``answer'' and ``correct'' in Figure \ref{fig:concept-tool}.
    \item \emph{Alternative} --- one of a set of concepts which together are universal or nearly so, but they do not appear together in the same name.
    To qualify, the words representing these concepts should not be synonyms: if they were, they would all describe a single concept.
    Rather, the alternatives may represent a difference of opinions about how to portray a certain attribute of the variable's intent.
    This could be based on different points of view (e.g.\ ``question'' vs.\ ``answer'' in Figure \ref{fig:concept-tool}), or else it can reflect different levels of generalization, where one concept is more general (e.g.\ ``data'') while another is more focused (e.g.\ ``scores'').
    In some cases one of the alternatives may be \emph{dominant} over the other alternatives (that is, universal in its own right).
    \item \emph{Optional} --- a substantial fraction of names include this concept, but others do not contain anything related.
    Example: the concept ``per'' in Figure \ref{fig:concept-tool}.
    This reflects a difference of opinions about whether something is worth including at all.
    \item \emph{Rare} --- a relevant concept that is used by only a small fraction of the names.
    This is a special case of an optional concept, where the prevailing consensus is that it is not worth using.
    Example: ``number'' in Figure \ref{fig:concept-tool}.
\end{itemize}
Note that this is slightly different from the classification used by Regev et al., which was based only on frequency and disregarded semantics \cite{regev21}.

A possible question is whether to keep correlated concepts separate: should we perhaps unite them into a single more specific concept?
For example, in Experiment 1 names given for the constant \code{GLOBAL\_A} always included the two concepts of ``index'' and ``contact info'' together.
This could be interpreted as reflecting the joint concept of ``location of contact info''.
However, upon reflection it seems better to retain the distinct concepts, as this facilitates a more precise analysis of concepts and the words that represent them.
This is especially true if there are any instances where not all the constituent concepts are indeed used together.

\section{Results on Name Use and Structure}
\label{sect:res1}

The following subsections describe the results of our reproduction of the study of Feitelson et al.\ \cite{feitelson22}.
The first two subsection answer Research Question \ref{rq:reproduce}, and Section \ref{sect:concepts-analysis} answers \ref{rq:concepts}.

\subsection{Name Reuse}

\begin{table}\centering
\caption{\label{tab:2hit}\sl
Results of name reuse for all the variables in both experiments.
\emph{$N$}: number of answers regarding this variable;
\emph{Diff}: number of different names given;
\emph{Divers=Diff/$N$}: diversity of names;
\emph{Max}: maximal answers giving the same name;
\emph{Focus=Max/$N$}: focus, the probability of the most popular name;
\emph{P2hit}: estimated probability of two participants using the same name.
Compare with Table 1 of \cite{feitelson22}.
}
\begin{tabular}{@{}llcccccc@{}}
\hline
\emph{Exp} & \emph{Variable} & \emph{$N$} & \emph{Diff} & \emph{Divers} & \emph{Max} & \emph{Focus} & \emph{P2hit} \\
\hline

exp1	& \code{GLOBAL\_A}	& 71	& 46	& 0.647	& 11	& 0.154	& 0.0450 \\

	& \code{GLOBAL\_B}	& 70	& 29	& 0.414	& 32	& 0.457	& 0.2265 \\

	& \code{GLOBAL\_C}	& 70	& 38	& 0.542	& 8	& 0.114	& 0.0506 \\

	& \code{GLOBAL\_D}	& 71	& 36	& 0.507	& 17	& 0.239	& 0.0890 \\

	& \code{a}	& 69	& 29	& 0.420	& 27	& 0.391	& 0.1837 \\

	& \code{b}	& 70	& 26	& 0.371	& 28	& 0.400	& 0.1832 \\

	& \code{c}	& 67	& 35	& 0.522	& 13	& 0.194	& 0.0701 \\

	& \code{d}	& 63	& 35	& 0.555	& 8	& 0.126	& 0.0531 \\

	& \code{e}	& 64	& 20	& 0.312	& 22	& 0.343	& 0.1552 \\

	& \code{f}	& 72	& 58	& 0.805	& 4	& 0.055	& 0.0231 \\

	& \code{g}	& 63	& 18	& 0.285	& 27	& 0.428	& 0.2149 \\

	& \code{h}	& 63	& 37	& 0.587	& 19	& 0.301	& 0.1080 \\

	& \code{m}	& 63	& 49	& 0.777	& 5	& 0.079	& 0.0284 \\

	& \code{n}	& 72	& 31	& 0.430	& 32	& 0.444	& 0.2129 \\

	& \code{o}	& 64	& 50	& 0.781	& 6	& 0.093	& 0.0307 \\

	& \code{p}	& 66	& 31	& 0.469	& 28	& 0.424	& 0.1942 \\

	& \code{q}	& 61	& 24	& 0.393	& 19	& 0.311	& 0.1308 \\

	& \code{r}	& 63	& 18	& 0.285	& 21	& 0.333	& 0.1751 \\

	& \code{s}	& 65	& 30	& 0.461	& 26	& 0.400	& 0.1791 \\

	& \code{t}	& 62	& 25	& 0.403	& 35	& 0.564	& 0.3277 \\

\hline

exp2f1	& \code{A}	& 45	& 32	& 0.711	& 4	& 0.088	& 0.0409 \\

	& \code{B}	& 45	& 36	& 0.800	& 4	& 0.088	& 0.0350 \\

	& \code{C}	& 43	& 32	& 0.744	& 8	& 0.186	& 0.0578 \\

	& \code{D}	& 44	& 27	& 0.613	& 7	& 0.159	& 0.0671 \\

	& \code{E}	& 42	& 37	& 0.880	& 3	& 0.071	& 0.0306 \\

exp2f2	& \code{A}	& 44	& 33	& 0.750	& 7	& 0.159	& 0.0526 \\

	& \code{B}	& 43	& 38	& 0.883	& 4	& 0.093	& 0.0319 \\

	& \code{C}	& 44	& 38	& 0.863	& 3	& 0.068	& 0.0309 \\

	& \code{D}	& 44	& 31	& 0.704	& 5	& 0.113	& 0.0464 \\

exp2f3	& \code{A}	& 44	& 33	& 0.750	& 3	& 0.068	& 0.0351 \\

	& \code{B}	& 43	& 36	& 0.837	& 3	& 0.069	& 0.0319 \\

	& \code{C}	& 44	& 24	& 0.545	& 11	& 0.250	& 0.1012 \\

	& \code{D}	& 44	& 35	& 0.795	& 5	& 0.113	& 0.0382 \\

	& \code{E}	& 44	& 21	& 0.477	& 18	& 0.409	& 0.1931 \\

	& \code{F}	& 43	& 31	& 0.720	& 8	& 0.186	& 0.0600 \\

\hline
\end{tabular}
\end{table}

We start with a reproduction of the analysis of name use as done in the original study of Feitelson et al.\ \cite{feitelson22}.
Table \ref{tab:2hit} summarizes the results regarding the variability of names, using the metrics of focus and diversity defined above.
It also shows the estimated probability that two developers use the same name (denoted $P2hit$), using the formula from the original study.

\begin{figure}\centering
\includegraphics[width=0.67\columnwidth]{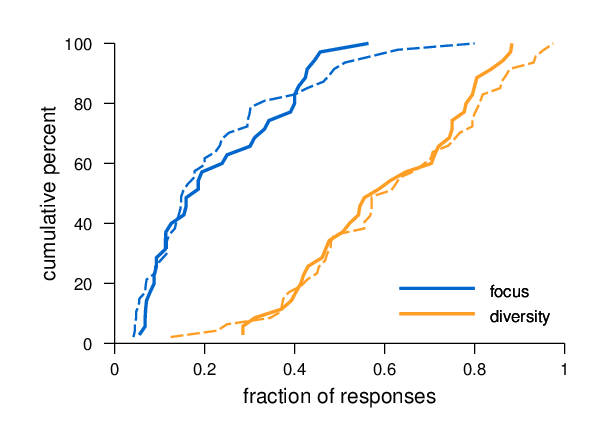}
\vspace*{-4mm}
\caption{\label{fig:focus-divers}\sl
Cumulative distribution functions of focus and diversity of names given to all variables in both experiments.
The dashed lines are the equivalent results from the original study (Figure 6 of \cite{feitelson22}).}
\end{figure}

Figure \ref{fig:focus-divers} shows the cumulative distribution functions (CDFs) of the diversity and focus results across all variables in both experiments.
They are amazingly similar to the CDFs found in the original study, indicating that the mix of variables with different levels of focus and diversity is similar in experiments performed at different times and using different methodologies.
The focus tends to be low, with the most popular name used in between 6--45\% of the responses.
The diversity tends to be higher, with 40--90\% of the names given by experiment participants being different from each other.

\begin{figure}\centering
\includegraphics[width=0.67\columnwidth]{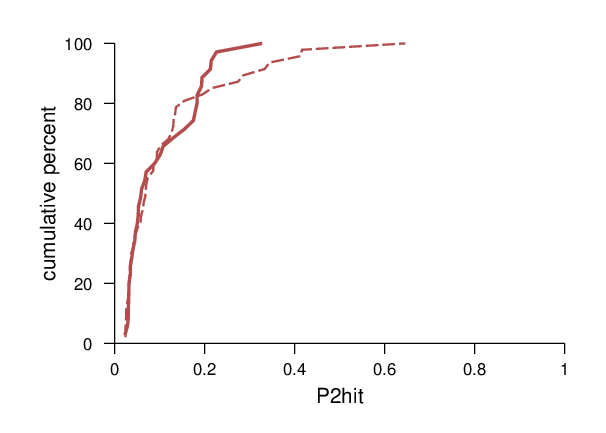}
\vspace*{-4mm}
\caption{\label{fig:p2hit-all}\sl
Cumulative distribution function of $P2hit$ for all variables in both experiments.
The dashed line is the equivalent result from the original study (Figure 8 of \cite{feitelson22}).}
\end{figure}

Figure \ref{fig:p2hit-all} shows the CDF of $P2hit$ results.
Again this is very close to the distribution obtained in the original study.
$P2hit$ tends to be low, and the median is only 6.0\% probability that two subjects would choose the same name.
The main divergence from the results of the original study is in the tail of the distribution.
In the original study there were two cases where a large fraction of the subjects chose exactly the same name.
These were cases where a single short word was the obvious choice.
In the present experiments there was only one such case, but with a much milder effect.

To summarize, our results in the current experiments closely match the results from the original study in terms of the distribution of names.

\subsection{Name Lengths}

\begin{figure}\centering
\includegraphics[width=0.67\columnwidth]{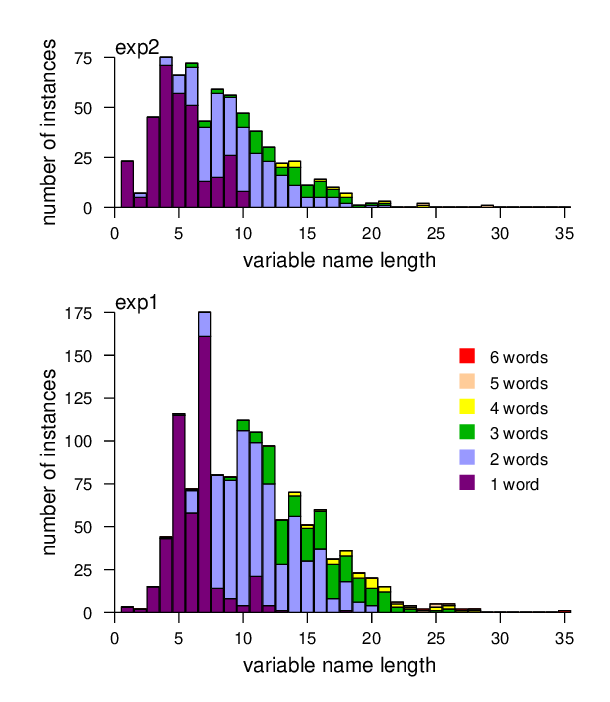}
\vspace*{-4mm}
\caption{\label{fig:var-len}\sl
Histograms of variable name lengths for all variables in both experiments, with partitioning into words.
Compare with Figure 4 of \cite{feitelson22}.}
\end{figure}

We also compared the distributions of lengths of names in the reproduction experiments with the distribution of name lengths in the original study.
The histograms of lengths are shown in Figure \ref{fig:var-len}.
The longest name seen was \code{potential\_winners\_contact\_and\_score\_list} (in Experiment 1), which is composed of 6 words and has 35 characters (excluding the underscores).

The distributions seen in Figure \ref{fig:var-len} are pretty similar to that of Figure 4 in the original study.
The differences are:
\begin{itemize}
    \item The names are generally shorter.
    In the original study the average length was 11.8 letters, and in the current experiments it was 9.8 letters.
    Comparing the histograms of the two experiments, it is also obvious that the names in Experiment 2 tended to be shorter than those in Experiment 1.
    Moreover, there were many more single-word names: 48\% in Experiment 2, 35\% in Experiment 1, but only 22\% in the original study.
    A possible explanation for these differences is that the reproduction experiments contained concrete and short codes.
    Experiment 2 in particular was composed of short independent functions.
    Thus the scope of the naming was much smaller, allowing participants to use shorter names.
    \item The distribution can be divided into single-word names and multi-word names.
    In the original study these formed two distinct modes, but in the reproduction they are adjacent.
    The single word names are mostly 3--7 letters long, wider than the 4--5 letters of the original study (which had an especially sharp peak of 4-letter names).
    We do not have a good explanation for this difference.    
    The multi-word names are mostly 8--16 letters long in both the original study and in the reproduction.
\end{itemize}

We note that the distribution of lengths in all these experiments (including the original study) are not identical to the distribution seen in real production code: real code has many more single-letter variable names \cite{beniamini17}.
These usually reflect various programming idioms, such as using \code{i} for loop indices.
As our experiments do not ask for names for such indices it is not surprising that we see very few single-letter names.
In the original study, which was based on descriptions of scenarios, there were only 2 such names.
In the replication experiments, which are based on actual code, there were 26 (after removing cases where the name was the same as the letter that represented the variable in the experiment).
Most were in Experiment 2, for example in a local variable within a list comprehension construct.

\begin{figure}\centering
\includegraphics[width=0.67\columnwidth]{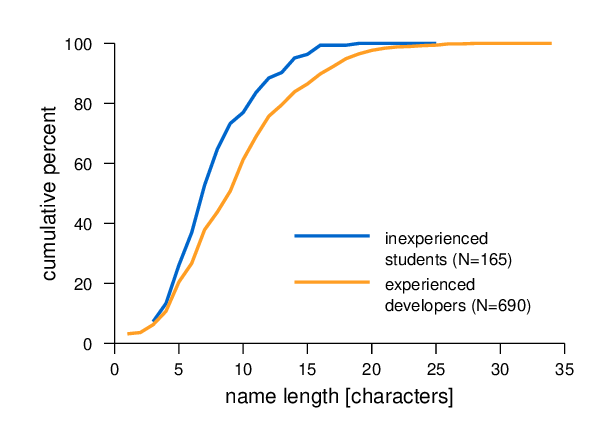}
\vspace*{-4mm}
\caption{\label{fig:exp-len}\sl
CDFs of variable name lengths for experienced developers as opposed to inexperienced students.
$N$ refers to the number of names given.
Compare with Figure 5 of \cite{feitelson22}.}
\end{figure}

Another interesting result obtained in the original study was an interaction between demographics and name length:
it was found that experienced developers tend to give slightly longer names than inexperienced students.
At the same time, it was also found that the distributions of name lengths produced by males and females were identical.
The same results were reproduced in the current experiment.
Experienced developers, defined as those with at least 5 years of experience, created names composed of 1.80 words on average, with an average length of 9.73 letters.
First degree students with at most 2 years experience created names composed of 1.62 words on average, and with an average length of 7.89 letters.
The cumulative distribution functions (CDFs) of lengths are shown in Figure \ref{fig:exp-len}.
A CDF that is lower and shifted to the right indicates a distribution of higher values.
As can easily be seen, the distribution representing experienced developers dominates that of students.
This result was found to depend on the identification of students --- inexperienced developers who were not students were practically indistinguishable from experienced ones.

\subsection{The Concepts in Names}
\label{sect:concepts-analysis}

Our main extension over the original study is in a deeper analysis of the concepts embedded in names, that is of the semantics of the words which compose the names.
This answers Research Question \ref{rq:concepts}.
It is based on identifying the concepts represented by the different words in the names, as described above in Section \ref{sect:concepts}.
We then drew a histogram of the frequency in which all the concepts in all the names appear.
This showed concentrations near the two ends --- very many concepts that appear only infrequently, and some that appear a lot.
Based on this we defined universal concepts to be those that appear in at least 85\% of the names, and rare ones to be those that appear in up to 15\%.
Concepts that appear in between 15\% and 85\% of the names are optional.

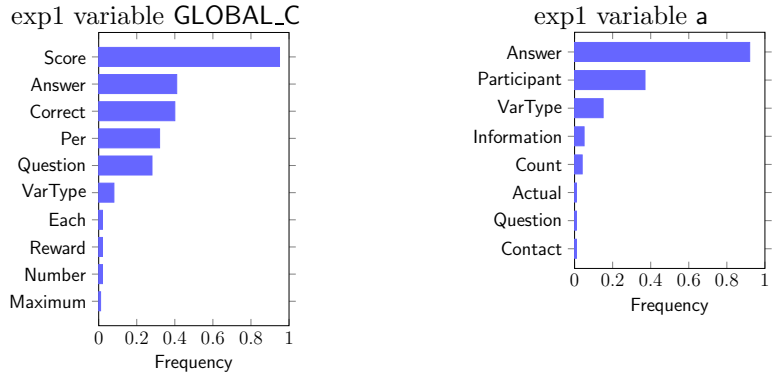
\begin{figure}\centering
  \parbox[t]{0.48\columnwidth}{\centering
    exp1 variable \code{GLOBAL\_C}

\begin{tikzpicture}[scale=0.6]
 \begin{axis}[
   font=\sffamily\large,
   width=120pt,
   height=185pt,
   scale only axis=true,
   xbar,
   bar width=12pt,
   ytick=data,
   yticklabel style={font=\sffamily\large},
  yticklabels={Maximum, Number, Reward, Each, VarType, Question, Per, Correct, Answer, Score},
   enlarge y limits=0.1,
   xlabel = {Frequency},
   xmin=0,xmax=1,
   xtick={0,0.2,0.4,0.6,0.8,1}]
   \addplot[fill=blue,color=blue!60,mark=none] coordinates { (0.01,1) (0.02,2) (0.02,3) (0.02,4) (0.08,5) (0.28,6) (0.32,7) (0.4,8) (0.41,9) (0.95,10)  };

 \end{axis}
\end{tikzpicture}
\vspace{2mm}
} \hfill \parbox[t]{0.48\columnwidth}{\centering
exp1 variable \code{a}

\begin{tikzpicture}[scale=0.6]
 \begin{axis}[
   font=\sffamily\large,
   width=120pt,
   height=149pt,
   scale only axis=true,
   xbar,
   bar width=12pt,
   ytick=data,
   yticklabel style={font=\sffamily\large},
  yticklabels={Contact, Question, Actual, Count, Information, VarType, Participant, Answer},
   enlarge y limits=0.1,
   xlabel = {Frequency},
   xmin=0,xmax=1,
   xtick={0,0.2,0.4,0.6,0.8,1}]
   \addplot[fill=blue,color=blue!60,mark=none] coordinates { (0.01,1) (0.01,2) (0.01,3) (0.04,4) (0.05,5) (0.15,6) (0.37,7) (0.92,8)  };

 \end{axis}
\end{tikzpicture}
}
\vspace*{-4mm}
    \caption{\label{fig:univ}\sl
    Example variables with varied concepts.}
\end{figure}

Figure \ref{fig:univ} shows two examples where there were concepts of many different types.
\code{GLOBAL\_C} in Experiment 1 is a constant representing the number of points given for each correct answer in the quiz.
Names given to this constant include \code{correct\_answer\_score}, \code{points\_per\_question}, and more (as was shown in Figure \ref{fig:concept-tool}).
The concept of ``score'', which also includes the word ``points'' and the abbreviation ``pts'', was universal: it appears in 95\% of the names.
Interestingly, some of the participants ascribed the score to the question, and others to the answer.
The concepts of ``question'' and ``answer'' where therefore alternatives, and never appeared together in the same name.
The concept of ``correct'' was optional --- many but not all participants specified that the points are specifically given to correct answers and not just any answer.
In addition there were a number of rare concepts, such as specifying that this is a ``number'' of points, or that its type is a ``constant''.

The other example is the variable \code{a}, which represents a single participant's quiz that is being processed.
The names given to this variable universally included the concept of ``answers'', which appeared in 92\% of the names.
There were two optional concepts describing these answers.
One indicated that the answers belong to a ``participant'' (37\% of the names).
The other indicated that the type was a ``list'' of answers (15\%).
A small number noted that the quiz also included ``information'', meaning the participant's email, and not only the answers to the questions.

\begin{table}\centering
    \caption{\label{tab:top-concepts}\sl
    Examples of universal concepts.
    The ones bellow the dividing line are prominent but not universal.
    \emph{Tot}: total names given for this variable;
    \emph{Top Con}: the most frequent concept;
    \emph{Next Con}: the next most common concept;
    $N$: names including each concept;
    \emph{Freq}: frequency of concept;
    \emph{Ratio}: of frequencies of top two concepts.}
    \begin{tabular}{@{}c@{ }c@{ }cl@{ }c@{ }cl@{ }c@{ }cc@{}}
    \hline
    \emph{Exp} & \emph{Var} & \emph{Tot} & \emph{Top Con.} & \emph{N} & \emph{Freq} & \emph{Next Con.} & \emph{N} & \emph{Freq} & \emph{Ratio} \\
    \hline
2f3 & \code{E} & 44 & Word	& 44	& 100\% & String	& 6	& 13\% & 7.73\\
1 & \code{G\_C} & 70 & Score	& 67	& 95\% & Answer	& 29	& 41\% & 2.32\\
1 & \code{p} & 66 & Place	& 63	& 95\% & Winner	& 14	& 21\% & 4.54\\
1 & \code{d} & 63 & Answer	& 59	& 93\% & VarType	& 38	& 60\% & 1.55\\
1 & \code{g} & 63 & Best	& 59	& 93\% & Score	& 55	& 87\% & 1.07\\
1 & \code{a} & 69 & Answer	& 64	& 92\% & Participant	& 26	& 37\% & 2.50\\
1 & \code{b} & 69 & Score	& 64	& 92\% & Total	& 16	& 23\% & 4.02\\
1 & \code{n} & 72 & Winner	& 66	& 91\% & VarType	& 8	& 11\% & 8.31\\
1 & \code{G\_B} & 70 & Answers	& 64	& 91\% & Correct	& 54	& 77\% & 1.18\\
1 & \code{e} & 64 & Contact	& 58	& 90\% & Participant	& 16	& 25\% & 3.62\\
1 & \code{r} & 63 & Name	& 57	& 90\% & Winner	& 18	& 28\% & 3.23\\
1 & \code{c} & 67 & Answer	& 60	& 89\% & VarType	& 37	& 55\% & 1.62\\
2f2 & \code{A} & 44 & Length	& 39	& 88\% & String	& 15	& 34\% & 2.60\\
1 & \code{t} & 62 & Winner	& 54	& 87\% & VarType	& 10	& 16\% & 5.46\\
2f3 & \code{C} & 44 & Separator	& 38	& 86\% & Char	& 8	& 18\% & 4.80\\
    \hline
1 & \code{s} & 65 & Message	& 54	& 83\% & VarType	& 13	& 20\% & 4.17\\
1 & \code{q} & 61 & Mail	& 49	& 80\% & Winner	& 16	& 26\% & 3.09\\
1 & \code{G\_D} & 71 & Number	& 56	& 78\% & Winner	& 54	& 76\% & 1.03\\
2f3 & \code{D} & 42 & Words	& 33	& 78\% & Titled	& 17	& 40\% & 1.96\\
2f3 & \code{B} & 43 & Case	& 33	& 76\% & First	& 21	& 48\% & 1.59\\
2f1 & \code{D} & 42 & Item	& 32	& 76\% & First	& 11	& 26\% & 2.94\\
1 & \code{h} & 63 & Participant	& 45	& 71\% & Result	& 31	& 49\% & 1.45\\
1 & \code{G\_A} & 71 & Number	& 53	& 74\% & Question	& 47	& 66\% & 1.12\\
2f3 & \code{A} & 44 & String	& 32	& 72\% & Input	& 13	& 29\% & 2.50\\
2f2 & \code{C} & 44 & Char	& 32	& 72\% & Random	& 20	& 45\% & 1.61\\
2f2 & \code{B} & 42 & Char	& 27	& 64\% & Pool	& 19	& 45\% & 1.43\\
1 & \code{o} & 64 & Index	& 40	& 62\% & Winner	& 39	& 60\% & 1.04\\
1 & \code{m} & 63 & Best	& 39	& 61\% & Participant	& 28	& 44\% & 1.39\\
2f1 & \code{A} & 42 & VarType	& 26	& 61\% & First	& 19	& 45\% & 1.36\\
2f1 & \code{C} & 42 & Result	& 25	& 59\% & Filtered	& 14	& 33\% & 1.80\\
2f1 & \code{B} & 45 & VarType	& 26	& 57\% & Filter	& 22	& 48\% & 1.19\\
2f2 & \code{D} & 44 & VarType	& 24	& 54\% & Output	& 21	& 47\% & 1.15\\
1 & \code{f} & 72 & Participant	& 35	& 48\% & Score	& 27	& 37\% & 1.31\\
2f3 & \code{F} & 43 & Output	& 21	& 48\% & String	& 21	& 48\% & 1.01\\
2f1 & \code{E} & 41 & Item	& 19	& 46\% & Include	& 11	& 26\% & 1.78\\
    \hline
    \end{tabular}
\end{table}

When analyzing names, the most prominent concepts are the \textbf{universal} ones.
But sometimes they stand out more than others.
The data for all variables in both experiments is shown in Table \ref{tab:top-concepts}.
The top part shows variable with concepts identified as universal.
In most cases the next most frequent concept appears with a significantly lower frequency.
But when there are no universal concepts, and especially when the top concept is not very frequent, there are usually other concepts with similar frequencies.

\begin{table}\centering
    \caption{\label{tab:pair-concepts}\sl
    Examples of pairs of concepts that nearly always appear together.
    \emph{Tot}: total names given for this variable;
    \emph{Con 1/2}: the concepts;
    $N$: names including each concept;
    \emph{Both}: the degree of overlap.}
    \begin{tabular}{@{}cccl@{ }cl@{ }cc}
    \hline
    \emph{Exp} & \emph{Var} & \emph{Tot} & \emph{Con.\ 1} & \emph{N} & \emph{Con.\ 2} & \emph{N} & \emph{Both} \\
    \hline
    1 & \code{GLOBAL\_A} & 71 & number & 53 & question & 47 & 42 \\
    1 & \code{GLOBAL\_B} & 71 & correct & 54 & answer & 64 & 53 \\
    1 & \code{GLOBAL\_C} & 71 & correct & 28 & answer & 29 & 23 \\
    1 & \code{GLOBAL\_D} & 71 & number & 58 & winner & 54 & 46 \\
    1 & \code{g} & 64 & best & 59 & score & 55 & 55 \\
    \hline
    \end{tabular}
\end{table}

As the examples in Figure \ref{fig:univ} show, there were two main types of relationships between concepts.
One was pairs of concepts that were strongly \textbf{correlated}, and nearly always appeared together.
These are simply cases where the desired semantics are best expressed by a phrase.
Examples we saw are listed in Table \ref{tab:pair-concepts}.
In some cases both members of the pair appeared in the majority of the names.
Note that even when the number of appearances of two concepts is very close, it does not necessarily mean that they always appear together.
It could be that there are similar numbers of instances where one appears and the other does not, in both directions.
A case in point is the variable \code{o}, where the concept ``winner'' appeared 39 times and the concept ``index'' appeared 40 times, but they appeared together only 24 times.
These concepts were therefore not considered correlated.

\begin{table}\centering
    \caption{\label{tab:alt-concepts}\sl
    Examples of alternative concepts used in names.
    \emph{Tot}: total names given for this variable;
    \emph{Alt}: the alternatives;
    $N$: names including each alternative.}
    \begin{tabular}{@{}cccl@{ }cl@{ }cl@{ }c@{}}
    \hline
    \emph{Exp} & \emph{Var} & \emph{Tot} & \emph{Alt. 1} & \emph{N} & \emph{Alt. 2} & \emph{N} & \emph{Alt. 3} & \emph{N} \\
    \hline
    1 & \code{G\_A} & 71 & question & 47 & part & 12 & answer & 4 \\
    1 & \code{G\_C} & 71 & answer & 29 & question & 20 & & \\
    1 & \code{G\_D} & 71 & winners & 54$^\dag$ & prizes & 7$^\dag$ & random & 4 \\
    1 & \code{f}    & 72 & score & 27 & results & 13 & answers & 6 \\
    1 & \code{b}    & 71 & score & 64 & results & 3 &  &  \\
    1 & \code{m}    & 63 & participant & 28 & winner/ & 23 &  &  \\
      &             &    &             &    & candidate & & & \\
    1 & \code{t}    & 63 & winner & 54$^\dag$ & chosen & 5$^\dag$ &  &  \\
    1 & \code{q}    & 63 & email & 49 & contact/ & 9 &  &  \\
      &             &    &       &    & info & & & \\
    2f1 & \code{A} & 45 & list & 26$^\ddag$ & values & 11$^\ddag$ &  &  \\
    2f1 & \code{C} & 45 & result & 25$^\dag$ & filtered & 14$^\dag$ &  &  \\
    2f2 & \code{B} & 45 & chars & 27$^\ddag$ & alphabet & 13$^\ddag$ &  &  \\
    2f2 & \code{C} & 45 & random & 20$^\ddag$ & list & 18$^\ddag$ &  &  \\
    2f2 & \code{D} & 45 & output & 21 & random & 18 &  &  \\
    2f3 & \code{F} & 44 & result & 21$^\dag$ & edited & 15$^\dag$ &  &  \\
    \hline
    \multicolumn{9}{l}{\footnotesize $^\dag$with one overlap; $^\ddag$with few overlaps.}
    \end{tabular}
\end{table}

The other common relationship was \textbf{alternatives}.
These are shown in Table \ref{tab:alt-concepts}.
The top one is in \code{GLOBAL\_A}, which represents the number of questions in the quiz.
This has a dominant concept representing ``number''.
Three alternative concepts describe what it is that the number represents.
The most commonly used denoted ``questions''.
A few others used an opposite concept denoting ``answers''.
And a third group used a concept of ``parts'' of a whole, meaning parts of the quiz.
Another example is \code{GLOBAL\_D}, where the dominant concept representing those who won was ``winners'', but in a few cases this was replaced by ``random'', which is how they were selected, or ``prize'', which is what they got.

The defining characteristic of alternatives is that they do not appear together in the same name.
However, there are isolated cases where this does happen.
For example, in \code{GLOBAL\_D} there was one instance of using the name \code{number\_of\_prize\_winners}.

As noted above, the \textbf{optional} concepts were often adjectives describing some other (possibly universal) concept.
In some cases they had a technical nature: specifying the type of the variable, e.g.\ that this is a ``list'', or being the ``current'' one.
In some cases inferior choices are made.
For example, the first function of Experiment 2 uses a \code{zip} of two lists, and the items were sometimes named ``first'' and ``second'' instead of indicating their true function.

Some of the \textbf{rare} concepts were even worse.
For example, the second parameter of the first function in Experiment 2 was a map to filter by.
Two respondents chose to call it \code{non\_empty}, which is indeed checked in the code, but obviously not its intent.
As another example, the variable denoting the ranking of winners for printing the notification message was most often named related to ``place'', but a couple just used ``string''.

finally, \code{GLOBAL\_A} also served to extract the email address of quiz respondents, which was included in the quiz results immediately after their answers to the quiz questions (so in the array of results the answers were in cells \code{0} through \code{GLOBAL\_A-1}, and the email was in cell \code{GLOBAL\_A}).
3 of the 71 participants gave names based on this use of the constant rather than on its use for the number of questions in the quiz.
This is an example of why it is a bad idea to use such dual-purpose constants.
It would have been better to create a separate constant for the email, which could be defined to be equal to the one reflecting the number of questions.

\begin{figure}\centering
\includegraphics[width=0.55\columnwidth]{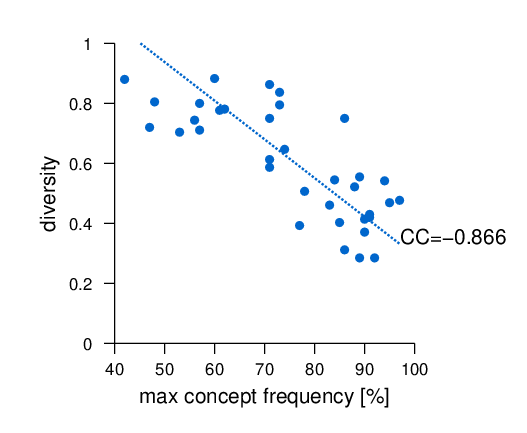}
\vspace*{-4mm}
\caption{\label{fig:freq-div}\sl
Relationship of maximal concept frequency and names diversity for all variables in both experiments.}
\end{figure}

Perhaps the most interesting variables are those with no clear pattern.
These exhibit a large diversity and many alternate or optional concepts, but no consensus regarding the main ones and how to express them.
This is demonstrated in Figure \ref{fig:freq-div}, which shows that diverse names are inversely correlated with having a dominant concept.
Characterizing them may therefore shed light on what causes difficulties in naming.


\subsection{Reasons for Name Choices}

\begin{table}\centering
    \caption{\label{tab:reasons}\sl
    Results of survey in Experiment 1 regarding reasons for name choices.
    $\mu$: the average of the answers.}

\vspace{1mm}
\noindent\begin{tabular}{@{}p{38mm}lc@{}}
\hline
\rule{0pt}{2ex}\raggedright ensure the names describe their purpose & $\mu=4.6$ &

\adjustbox{valign=t}{\begin{tikzpicture}[scale=0.5]
 \begin{axis}[
   font=\sffamily\LARGE,
   width=120pt,
   height=80pt,
   scale only axis=true,
   ybar,
   bar width=12pt,
   xtick=data,
   ylabel = {Percents},
   ymin=0,height=100 pt,
   ymax=80,
   ytick={0,20,40,60,80}
  ]
  \addplot[fill=blue,color=blue!60,mark=none] coordinates {(1,1.40845070422535) (2,1.40845070422535) (3,4.22535211267606) (4,18.3098591549296) (5,74.6478873239437) };
 \end{axis}
\end{tikzpicture}}
\\[1mm]
\rule{0pt}{2ex}\raggedright make the code understandable & $\mu=4.6$ &

\adjustbox{valign=t}{\begin{tikzpicture}[scale=0.5]
 \begin{axis}[
   font=\sffamily\LARGE,
   width=120pt,
   height=80pt,
   scale only axis=true,
   ybar,
   bar width=12pt,
   xtick=data,
   ylabel = {Percents},
   ymin=0,height=100 pt,
   ymax=80,
   ytick={0,20,40,60,80}
  ]
  \addplot[fill=blue,color=blue!60,mark=none] coordinates {(1,1.40845070422535) (2,1.40845070422535) (3,2.8169014084507) (4,22.5352112676056) (5,71.830985915493) };
 \end{axis}
\end{tikzpicture}}
\\[1mm]
\rule{0pt}{2ex}\raggedright keep the names as task-specific as possible & $\mu=4.0$ &

\adjustbox{valign=t}{\begin{tikzpicture}[scale=0.5]
 \begin{axis}[
   font=\sffamily\LARGE,
   width=120pt,
   height=80pt,
   scale only axis=true,
   ybar,
   bar width=12pt,
   xtick=data,
   ylabel = {Percents},
   ymin=0,height=62.5 pt,
   ymax=50,
   ytick={0,20,40}
  ]
  \addplot[fill=blue,color=blue!60,mark=none] coordinates {(1,2.85714285714286) (2,4.28571428571429) (3,12.8571428571429) (4,47.1428571428571) (5,32.8571428571429) };
 \end{axis}
\end{tikzpicture}}
\\[1mm]
\rule{0pt}{2ex}\raggedright make the names distinct from each other & $\mu=3.8$ &

\adjustbox{valign=t}{\begin{tikzpicture}[scale=0.5]
 \begin{axis}[
   font=\sffamily\LARGE,
   width=120pt,
   height=80pt,
   scale only axis=true,
   ybar,
   bar width=12pt,
   xtick=data,
   ylabel = {Percents},
   ymin=0,height=62.5 pt,
   ymax=50,
   ytick={0,20,40}
  ]
  \addplot[fill=blue,color=blue!60,mark=none] coordinates {(1,1.42857142857143) (2,8.57142857142857) (3,25.7142857142857) (4,41.4285714285714) (5,22.8571428571429) };
 \end{axis}
\end{tikzpicture}}
\\[1mm]
\rule{0pt}{2ex}\raggedright include all the relevant details in each name & $\mu=3.6$ &

\adjustbox{valign=t}{\begin{tikzpicture}[scale=0.5]
 \begin{axis}[
   font=\sffamily\LARGE,
   width=120pt,
   height=80pt,
   scale only axis=true,
   ybar,
   bar width=12pt,
   xtick=data,
   ylabel = {Percents},
   ymin=0,height=50 pt,
   ymax=40,
   ytick={0,20,40}
  ]
  \addplot[fill=blue,color=blue!60,mark=none] coordinates {(1,4.22535211267606) (2,18.3098591549296) (3,18.3098591549296) (4,33.8028169014084) (5,25.3521126760563) };
 \end{axis}
\end{tikzpicture}}
\\[1mm]
\rule{0pt}{2ex}\raggedright make the names good English sentences or expressions & $\mu=3.4$ &

\adjustbox{valign=t}{\begin{tikzpicture}[scale=0.5]
 \begin{axis}[
   font=\sffamily\LARGE,
   width=120pt,
   height=80pt,
   scale only axis=true,
   ybar,
   bar width=12pt,
   xtick=data,
   ylabel = {Percents},
   ymin=0,height=50 pt,
   ymax=40,
   ytick={0,20,40}
  ]
  \addplot[fill=blue,color=blue!60,mark=none] coordinates {(1,8.45070422535211) (2,15.4929577464789) (3,25.3521126760563) (4,32.3943661971831) (5,18.3098591549296) };
 \end{axis}
\end{tikzpicture}}
\\[1mm]
\rule{0pt}{2ex}\raggedright keep the names concise (avoid extra words) & $\mu=3.3$ &

\adjustbox{valign=t}{\begin{tikzpicture}[scale=0.5]
 \begin{axis}[
   font=\sffamily\LARGE,
   width=120pt,
   height=80pt,
   scale only axis=true,
   ybar,
   bar width=12pt,
   xtick=data,
   ylabel = {Percents},
   ymin=0,height=50 pt,
   ymax=40,
   ytick={0,20,40}
  ]
  \addplot[fill=blue,color=blue!60,mark=none] coordinates {(1,7.04225352112676) (2,15.4929577464789) (3,30.9859154929577) (4,30.9859154929577) (5,15.4929577464789) };
 \end{axis}
\end{tikzpicture}}
\\[1mm]
\rule{0pt}{2ex}\raggedright keep the names as general as possible & $\mu=2.6$ &

\adjustbox{valign=t}{\begin{tikzpicture}[scale=0.5]
 \begin{axis}[
   font=\sffamily\LARGE,
   width=120pt,
   height=80pt,
   scale only axis=true,
   ybar,
   bar width=12pt,
   xtick=data,
   ylabel = {Percents},
   ymin=0,height=50 pt,
   ymax=40,
   ytick={0,20,40}
  ]
  \addplot[fill=blue,color=blue!60,mark=none] coordinates {(1,14.0845070422535) (2,39.4366197183099) (3,26.7605633802817) (4,12.6760563380282) (5,7.04225352112676) };
 \end{axis}
\end{tikzpicture}}
\\[1mm]
\rule{0pt}{2ex}\raggedright include an indication of the variable's type & $\mu=2.5$ &

\adjustbox{valign=t}{\begin{tikzpicture}[scale=0.5]
 \begin{axis}[
   font=\sffamily\LARGE,
   width=120pt,
   height=80pt,
   scale only axis=true,
   ybar,
   bar width=12pt,
   xtick=data,
   ylabel = {Percents},
   ymin=0,height=50 pt,
   ymax=40,
   ytick={0,20,40}
  ]
  \addplot[fill=blue,color=blue!60,mark=none] coordinates {(1,23.943661971831) (2,32.3943661971831) (3,21.1267605633803) (4,19.7183098591549) (5,2.8169014084507) };
 \end{axis}
\end{tikzpicture}}
\\
\hline

\end{tabular}
\vspace{2mm}

\end{table}

Another extension over the original study was  a short survey regarding the considerations used in naming, that was included in Experiment 1.
Participants were presented with a series of statements, and were asked to register their agreement on a 5-point scale.
The results of this survey are shown in Table \ref{tab:reasons}.
The two top reasons cited where that names should describe their purpose and promote the understandability of the code.
Keeping names concise by avoiding extra words received a relatively low score.
It therefore seems that developers prefer descriptiveness over brevity.

\section{Reproduction Experiment 3}
\label{sect:exp3}

Our third experiment reproduces the second experiment of the original study, where participants were instructed on using the 3-step model to construct names.
As one of the results was that the names they created were longer than when they did not receive such instruction, we added another treatment in which we suggested to participants that long names are better.
This enables a check whether using the model explicitly actually makes a difference, or perhaps similar results may be achieved with a simpler intervention --- answering Research Question \ref{rq:better}.

\subsection{Experimental Materials}

The experimental materials were based on 5 independent programming scenarios, similar to those used in the original study.
Each scenario presented a programming problem.
As in the original study, we attempted to create non-trivial but general problems that any programmer can understand and relate to.
Two of the scenarios were taken from the original study: the minesweeper game and the mouse looking for cheese in a maze.
Three scenarios were new: a library system, a four-in-a-row game, and a friend-bring-friend campaign in a high-tech company.
In each case the description of the scenario was followed by 3--5 questions about names of variables that may be expected to appear in a program that solves this problem.
For example, in the library scenario participants were asked to name the constant specifying the maximal number of books that a subscriber may loan, the Boolean field in a \code{book} object specifying whether the book is currently on loan, and the data structure holding the library's subscribers.
There were a total of 19 such questions in all 5 scenarios.
No actual code was presented and no code was written.

The descriptions of the scenarios were given in Hebrew, as was done in the original study.
As shown there, this reduces the accessibility bias and reduces the risk that participants will be influenced by the wording used in the descriptions \cite{feitelson22}.

\subsection{Experiment Structure}

The experiment was composed of three sections.
The first section introduced the experiment as dealing with code comprehension.
It asked participants to provide their initial ideas and not look for ``correct'' answers.
It also stated that no personal data will be collected and that the results will be used for research purposes only, that the experiment is expected to take around 10 minutes, and that continuing to the questionnaire constitutes consent to participate.

The second section contained demographic background questions.
We asked about participants' sex, age, years of programming experience, and whether they are students.
Those answering that they are students were also asked about their degree and study year.

The final section of the experiment contained the 5 scenarios and the naming questions about each one.
This section started with an explanation that several programming scenarios will be presented in the following pages, and that participants are requested to give names to variables that take part in solutions to these scenarios.
The difference between the three treatments was limited to additional details that appeared after this explanation.
In one treatment nothing was added beyond the above statement. 
The second treatment noted that naming is important, and that research has shown that long detailed names which explain the role of the variable make the code more understandable.
The third treatment presented the 3-step model of naming, with an example (the same example about the display of the score in a bowling alley as in the original study), and asked participants to use this model when giving names.

The 5 scenarios and their questions followed.
In the third treatment a short one-line reminder of the 3-step model was shown at the top of each scenario.

\subsection{Experiment Execution}

This experiment was conducted online using the Qualtrics platform.
Participants were recruited via personal contacts and solicitation in our department’s computer labs.
As added motivation they were invited to enter a raffle for a 200 NIS gift card.
There were 150 participants in total, divided equally among the three treatments.
59\% were male and 41\% female.
Their programming experience (excluding studies) ranged from 0 to 28 years, with 37\% having 0--2 years, another 37\% with 3--4 years, and 24\% with 5 years or more.
In this experiment 91 of the participants (61\%) were active students, of which 62 where BSc students in their 3rd or 4th year, and 9 were graduate students.

The treatment shown to each participant was chosen at random.
All participants received the same 5 scenarios, but in a random order, so as to reduce threats related to learning and fatigue.

\section{Judging Name Quality}
\label{sect:judge}

The goal of this experiment was to see if the interventions --- being instructed about the 3-step naming model, or that longer names are better --- lead to the creation of better names.
Note that when we speak of name quality we do not mean whether they adhere to style conventions \cite{butler10,butler15}.
Rather, we are interested in name quality in terms of their semantics in the context of the different scenarios.
To assess this we recruited 6 external judges, who were not involved in the experiment in any way, and devised two alternative protocols for grading and ranking the names.
Each protocol was executed by 3 judges.

The judges were 4 males and 2 females, with an average of 7.8 years of programming experience (well above the average of the participants in the experiment).
They were paid 500 NIS for their work (approximately \$150).
They were assigned at random to either of the two judging protocols.
The judges assigned to each protocol all received exactly the same tasks, but they worked individually, without being exposed to what the other judges were doing.

\subsection{Name Normalization}

The first step in the analysis was to normalize the names.
The normalization consisted of converting them to all lowercase, with words separated by underscores (that is, to snake\_case format).
However, word separation was only applied when the participant separated the words.
For example, if the participant wrote \code{boardsize} as one word, it was not separated into \code{board\_size}.

Following the normalization all repetitions were removed.
This left us with 3 pools of unique names for each variable in each scenario, containing the names given under the 3 treatments.
Our goal is to find whether there exists a difference in quality between the names in these 3 pools, and if so, which treatment led to the best names.

A possible threat to the comparison between names is that some names are very similar to each other.
Such small differences can have little if any semantic meaning, and therefore may not reflect a real difference in quality.
We therefore tagged names that had a Levenshtein distance of 2 or less, meaning that one name could be transformed into the other by changing up to 2 letters.
Such tagged pairs were not compared to each other in the data presented to the judges.

\subsection{Ranking Protocol}

The ranking protocol was an extension of the judging protocol used in the original study.
In that study judges were presented with pairs of names, one from each treatment, and asked to judge which they thought was the better name in the context of the given programming scenario.
In the current experiment judges were presented with triplets of names, one from each treatment.

The data was presented to the judges in the form of an Excel workbook.
The workbook contained 19 sheets, one for each variable from one of the scenarios.
Each sheet started with the description of a scenario and a question about a variable.
These description and question were exactly the same as those shown to participants in the experiment.
Following was a table with 4 columns and 40 rows.
In each row, the first 3 columns contained different names given to this variable by participants in the 3 different treatments.
The judges were asked to decide which of the 3 names is the best, in the sense that it provides the best aid to understand the variable's role, and to copy it to the 4th column.
The tables shown to the 3 judges contained exactly the same 40 triplets of names, but the order of the rows was randomized differently for each judge.

This protocol is based on a direct comparison of names given under the 3 treatments.
The treatment that leads to higher quality names is identified as the one whose names were most often selected as the best names.

\subsection{Grading Protocol}

The second judgment protocol is based on grading individual names, rather than on comparing names to each other.

The data was again presented to the judges in the form of an Excel workbook, with 19 sheets, one for each variable from one of the scenarios.
Each sheet again started with the description of a scenario and a question about a variable, as in the previous judging protocol.
Following was a table, where each row contained one variable name from one of the treatments.
All the names given under all of the treatments were included, so the number of rows was different in each case.
The names were presented in a random order, which was different for each of the 3 judges.
The judges were asked to give a grade to each name, on a scale of 1 to 10.
1 meant that the name does not help at all to understand the variable's role, and 10 meant that if helps very much to understand the variable's role.

This protocol assesses the quality of each name individually, without any direct comparisons between them.
The treatment that leads to higher quality names is identified as the one whose names were given higher grades.

\section{Name Quality Results}
\label{sect:res2}

The 3 judges who performed the ranking protocol needed to select the best name of 3 from 40 triplets of candidates from 19 naming questions, for a total of 760 triplets.
One judge did not provide an answer in one case, so we were left with 759 results.
In 477 of them (which are 62.8\%) all 3 judges agreed on the best name.
In an additional 277 (36.4\%) 2 judges agreed.
So in total a majority of the judges (2 or more) agreed in a whopping 99.3\% of the cases.
The analysis focused on these agreed names.

Assigning the chosen names to treatments is complicated by the fact that some names were actually used by participants in more than one treatment.
Specifically, 60.9\% of the chosen names could be uniquely assigned to a single treatment, but 20.4\% appeared in 2 treatments, and 18.6\% were used in all 3 treatments.
In each triplet of candidates such names represented one specific treatment, but in principle they could also have represented another treatment.
We therefore assigned scores inversely to the number of treatments in which a name appeared.
Thus chosen names that were used in only one treatment gave one point to that treatment.
Chosen names that were used in 2 treatment gave $\frac 1 2$ a point to each of these treatments, and names that were used in all 3 gave $\frac 1 3$ of a point to each.

With this accounting, names representing the 3-step model treatment were chosen as the best in 41.3\% of the cases.
Names representing the treatment where participants were told that long detailed names are better were chosen in 30.9\% of the cases.
And names representing the treatment where participants received no specific instructions were chosen in the remaining 27.7\% of the cases.
Thus the 3-step model treatment names were chosen more often than names from either of the other two treatments, but not a majority of the time.

\begin{figure}\centering
\includegraphics[width=0.67\columnwidth]{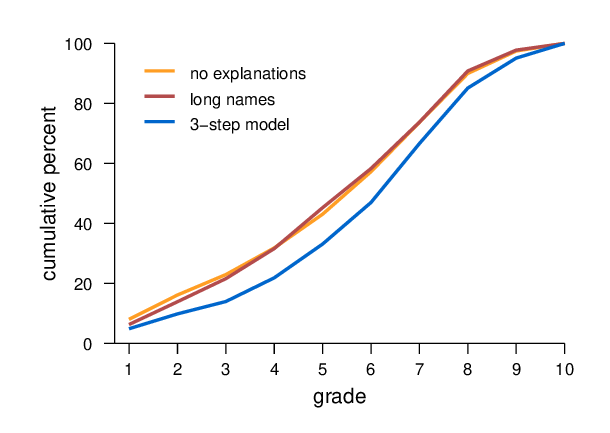}
\vspace*{-4mm}
\caption{\label{fig:grades}\sl
CDFs of grades given by the judges to names generated under the three treatments.
Grades given to names generated by participants who were coached about the 3-step model for name formation tend to be higher.}
\end{figure}

The results of the grading protocol are shown in Figure \ref{fig:grades}.
This graph shows CDFs of the grades given to names from the 3 treatments.
It is easily seen that two of the treatments --- where participants received no specific instructions, and where they were told that long detailed names are better --- produced essentially the same distribution of grades.
The third treatment, where participants were instructed about the 3-step model for name formation, led to a distribution that is shifted to higher values.
The average grades in the first two distributions were 5.59 and 5.60; in the third it was 6.22 --- 11\% higher.

\begin{figure*}\centering
\includegraphics[width=0.67\columnwidth]{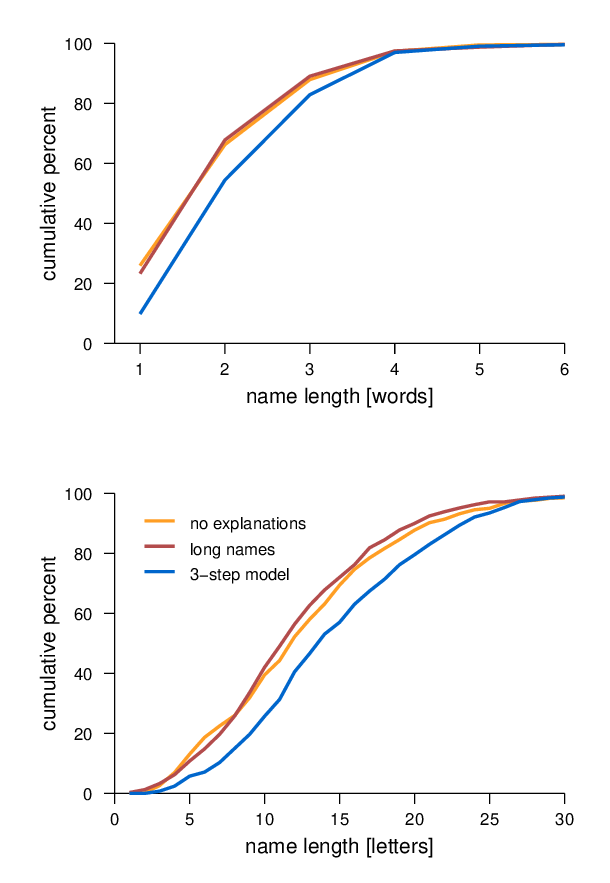}
\vspace*{-4mm}
\caption{\label{fig:lengths}\sl
CDFs of variable name lengths for names generated under the three treatments.
Names generated by participants who were coached about the 3-step model for name formation tend to be longer.}
\end{figure*}

Another result from the original study was that judges tended to prefer longer and more detailed names.
Given the grading of names which we performed here (and did not exist in the original study) we can analyze this in two steps.
First, Figure \ref{fig:lengths} shows the distribution of name lengths for names given under the 3 treatments.
The results are essentially similar to the results regarding the grades.
In the first two treatments the distributions of lengths are essentially the same.
In the 3-step model treatment the distributions are shifted to longer lengths, both when length is quantified using the number of letters and when it is quantified using the number of words in the name.
For example, when given no specific instructions and when instructed about using long detailed names, 26.0\% and 23.3\% of the names that the participants produced, respectively, contained only one word.
In the third treatment, where they were instructed about the 3-step model, only 9.8\% of the names had only one word.
The average lengths in the first 2 treatments were 12.9 and 12.5 letters and 2.23 and 2.24 words.
In the third treatment they were 15.0 letters and 2.57 words: 2.1--2.5 letters longer and 0.33 words longer.

\begin{figure}\centering
\includegraphics[width=0.63\columnwidth]{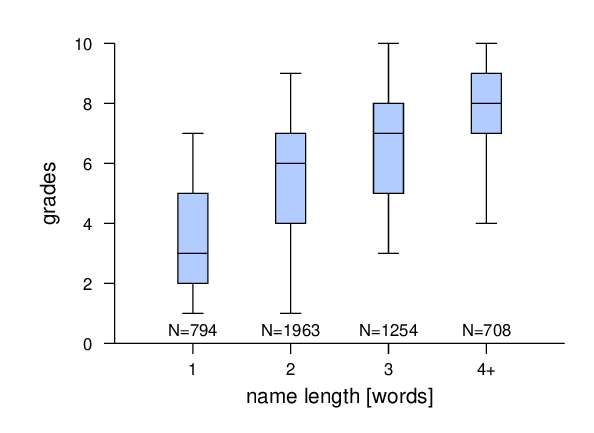}
\vspace*{-4mm}
\caption{\label{fig:len-grades}\sl
Distributions of grades for names with different numbers of words.}
\end{figure}

The above results show that the 3-step model treatment produced longer names and names that received higher grades.
But they do not show that these two results are related.
The second step of the analysis fills this gap by looking at the distribution of grades given to names with different lengths, from all three treatments together.
The results are shown in Figure \ref{fig:len-grades}.
Obviously grades correlate with length, and names with more words receive higher grades.
The correlation coefficient is 0.49, and performing a linear regression indicates that each additional word increases the grade by 1.14 points on average.

Together these analyses reproduce and add details to the results of the original study.
We reproduce the result that instructing participants about the 3-step model for naming facilitates the creation of better names.
We extend this by also showing that a simpler intervention, where participants are only instructed that longer more detailed names are better, does not produce such an effect.
We also reproduce the result that using the model leads to the creation of longer names, and directly show that longer names receive higher grades when judged on how much they help to understand a variable's role.

The implication for practice is that coaching about the 3-step model may be expected to help developers to create better names.
Just advising them that longer detailed names are better is not expected to deliver improved performance.
Likewise, mechanical aids such as IDE plugins that check name lengths will probably not work.
The 3-step model works because it guides developers in how to think about the names.

\section{Threats to Validity}
\label{sect:threats}

The experiments reported here have an exploratory nature, rather than trying to establish causality.
Therefore the main threats to validity concern construct validity and external validity.
More generally, reproduction is a major vehicle for addressing threats to validity.
In this sense all our current experiments are intended to verify the validity of the observations made in the original study of Feitelson et al.\ \cite{feitelson22}.
In addition, we use independent complementary experiments, so analogous results can further reduce threats.
For example, Experiment 1 may suffer from threats to internal validity due to writing experiment-specific code and influencing names via functions names.
But these threats do not exist in Experiment 2.
So if the results of Experiment 1 match those of Experiment 2, this is evidence that the threats did not materialize.

One source of threats to construct validity is that an experiment is not similar to real-life coding.
Participants may invest less effort, as witnessed by some superficial names they gave (reflecting technicalities rather than substance).
Such names are probably less common in real development.
Also, these experiments employ relatively short blocks of code, which justify using relatively short names.
In real code names can be much longer \cite{amit22}.
We therefore need to perform experiments in such settings too.
Another threat is that maybe some participants happened to take part in more than one of the experiments, and their experience with one could affect their behavior in another.
We can not know for sure, as participation in the experiments was anonymous.

One of our goals was to avoid an accessibility bias by using a general Hebrew description, and renaming in actual code rather than specific descriptions of variables.
But the function names in Experiment 1 could compromise this.
For example, one helper function was named \code{process\_the\_answers\_of\_one\_participant}, and the word \code{participant} figured prominently in names referring to the people who participated in the quiz.
However, the word \code{answer} was not used much, and the word \code{question} was very dominant, despite not having been mentioned by us.
Likewise, regarding the helper function \code{select\_who\_get\_prizes}, the word \code{prize} was seldom used, while the word \code{winner} was omnipresent.
These observations testify that our study participants understood the code and used their own words, and that accessibility was probably not a decisive factor.

Our analysis of the concepts in names is based on personal judgment.
For example, in the second function of Experiment 2, the parameter is the length of the random string to generate.
Names given to this parameter often included ``length'' or ``number'', but never both together.
In this context we decided to consider them as representations of the same concept.
But separating them into different concepts and considering them to be alternatives is also possible.
We reduced this threat by having two authors reach a consensus about the concepts, but independent analyses may reach different conclusions.
The best approach to validate our results is for other research groups to replicate these experiments using their own experimental materials and their own analysis.

Likewise, the analysis of name quality in the third experiment is also based on judgment.
We mitigated the risks involved by employing multiple external judges who worked independently, and by using two alternative judgment protocols.

As for external validity, our experimental subjects are in general not native English speakers.
Hence they may fashion names that may sound jarring to native English speakers, for example \code{won\_participants} rather than \code{winning\_participants}.
They may also use linguistic anti-patterns like a singular name for an array \cite{arnaoudova16}.
This affects the distributions we see, and causes them to be more diverse, because more variants of words are used, and perhaps also more questionable concepts.
Note, however, that the majority of software developers in the world are not native English speakers, so our results may actually be generally representative.
However, comparing the names produced by native English speakers and non-native-English speakers is an interesting avenue for further research.

\section{Conclusions}
\label{sect:conc}

We reproduced many of the results from Feitelson et al.\ \cite{feitelson22}, as summarized in Table \ref{tab:rep}, thereby answering Research Question \ref{rq:reproduce}.
Importantly, our reproduction used alternative experimental methodologies, notably using renaming of variable names in code with meaningless names, as opposed to giving names based on a verbal description of a scenario.
This corroborates the validity of the results beyond the use of an exact replication.

\begin{table*}\centering
    \caption{\label{tab:rep}\sl
    Reproduced results from Feitelson et al.\ \cite{feitelson22}.}
    \begin{tabular}{@{}p{50mm}cp{50mm}@{}}
    \hline
    \emph{Result} & \emph{Rep?} & \emph{Details} \\
    \hline
    Distribution of name lengths & partly &
         names were shorter \\
     & & distribution was not bimodal \\[1mm]
    Division of names to words & partly &
         \parbox[t]{50mm}{\raggedright there were more single-word names} \\[5mm]
    \parbox[t]{50mm}{\raggedright Experienced developers give longer names} & yes &
         \parbox[t]{50mm}{\raggedright inexperienced need to be students} \\[5mm]
    Sex does not affect name lengths & yes & \\[1mm]
    Distribution of focus of names & yes & \\[1mm]
    \parbox[t]{50mm}{\raggedright Distribution of diversity of names} & yes & \\[5mm]
    \parbox[t]{50mm}{\raggedright Distribution of probability to give same name} & yes & \\[5mm]
    Judges tend to agree which name is better & yes &
         \parbox[t]{50mm}{\raggedright near universal agreement of majority} \\[5mm]
    Judges prefer names from 3-step model & partly &
         \parbox[t]{50mm}{\raggedright conditions were different (3 choices rather than 2)} \\[5mm]
    Judges prefer longer names & yes &
         based on correlation with grades \\
    \hline
    \end{tabular}
\end{table*}

We also performed several extensions to the experiments and analyses of the original study.
For example, we suggest a more detailed classification of concepts based on their frequency and their relationship to each other.
This can be used to explain the sources of naming variations (Research Question \ref{rq:concepts}).
We also add a third treatment and a new judging methodology to the intervention experiment.
This shows that merely suggesting that longer names are better does not affect naming behavior, and that instruction about the 3-step model is required to generate names that receive higher grades (answering Research Question \ref{rq:better}).

The goal behind these experiments and analyses is to better understand how developers choose names, and when and why naming is difficult.
Our results point where we should look for answers:
we should focus our research on variables that receive long and diverse names.
Long names imply using a multiplicity of concepts, and specifically many optional and rare concepts rather than few universal ones.
Diversity, especially as a result of a lack of a clear pattern of dominant concepts, implies uncertainty and conflicting opinions on what the names should be and what they should include.
Together these characteristics identify naming pain-points that stand to benefit from better understanding and from the development of mitigation strategies.

The same considerations apply to putting these results to practical use.
Developers today are required to ``use meaningful names'', but without any coaching on how to achieve this objective.
Our results suggest a possible approach to train developers to think effectively about naming.
Start by assembling a sizable group of developers, say 20 to 30, and present them with a naming challenge like our Experiment 1.
Select names that are expected (based on the results reported above) to exhibit high diversity, and analyze them interactively, showing the participants how the names they had chosen can be dissected into sets of concepts, and the differences between the chosen concepts and ways to express them.
This can lead to a discussion of the considerations involved.
Why did some of the developers in the group choose to include or exclude a certain optional concept?
What are the implications of using one or another alternative concept?
Is there any value in adding rare concepts?
Point out warning signals, such as concepts that describe technical issues, which indicate superficial naming and a need for improvement.
Finally, introduce the 3-step model for forming names: selecting the concepts to include, the words to represent them, and the order of these words.
Such a training session can equip developers with guidance on how to think about naming, and inform them on how others see these issues, thereby improving the use of names to convey information.

\section*{Experimental Materials}

Complete experimental materials and results are available on Zenodo at
DOI 10.5281/zenodo.10665377.


\bibliographystyle{myabbrv}
\bibliography{abbrv,se,misc}

\end{document}